\documentclass[twocolumn,showpacs,preprintnumbers,amsmath,amssymb,showkeys]{revtex4}

\usepackage{epsfig}
\usepackage[utf8]{inputenc}
\usepackage[T1]{fontenc}

\usepackage{graphicx}
\usepackage{dcolumn}
\usepackage{bm}
\usepackage{color}
\usepackage{slashed}
\usepackage{amsmath}
\usepackage[abs]{overpic}
\usepackage[section]{placeins} 
\usepackage{mathtools}

\newcommand\eqn[1]{(\ref{#1})}

\begin{document}

\title{On the contribution of different coupling constants in the infrared regime of Yang-Mills theory: a Curci-Ferrari approach.}

\author{Mat\'ias Fern\'andez$^a$}

\author{Marcela Pel\'aez$^{a}$}

\affiliation{\vspace{.2cm}
$^a$Instituto de F\'{\i}sica, Facultad de Ingenier\'{\i}a, Universidad de la Rep\'ublica,\\
J.~H. y Reissig 565, 11000 Montevideo, Uruguay.}

\date{\today}

\begin{abstract}
We investigate the influence of the different vertices of two-point correlation functions in the infrared regime
of Yang-Mills theory using a phenomenological description.
This regime is studied in Landau-gauge and using perturbation theory within a phenomenological massive model.
We perform a one-loop calculation for two-point correlation functions taking into
account the different role of the various interactions in the infrared. Our results show a good agreement with the
lattice data. 
\end{abstract}


\maketitle

\section{Introduction}

Several efforts have been made in order to understand the infrared (IR) behavior of 
correlation functions in Yang-Mills theories. The IR is usually known as the non-perturbative
regime since standard perturbation theory based on Faddeev-Popov Lagrangian is no longer valid.
For this reason other semi-analytical techniques have been implemented (see e.g. \cite{Pawlowski:2003hq,Alkofer:2000wg,PhysRevD.65.094039,PhysRevD.67.094020,Bloch03,Fischer:2004uk,NATALE2007,Fischer:2008uz,RodriguezQuintero10,Huber:2012kd,PhysRevD.78.025010,refId0,AgPa08,Alkofer:2000wg,Aguilar:2004sw,Wetterich92,Berges:2000ew}).
Correlation functions are important since they can be related to scattering amplitudes and used as a testing ground for different approaches in the
study of IR Yang-Mills or QCD. 
However, these quantities are gauge dependent and it is therefore necessary to compare analytic results with gauge fixed simulations.

At present, one of the main approaches to nonperturbative problems is the lattice simulation.
In the last decades Landau gauge-fixed lattice calculations have improved considerably and a consensus has been reached on several unexpected
features.
First, they found that gluon two-point correlation function behaves as a massive propagator in the IR \cite{Cucchieri_08b,Cucchieri_09,Cucchieri_08,Bogolubsky:2009dc,Dudal10}.
Second, that the coupling constants remain moderate even at low momenta \cite{Bogolubsky:2009dc,PhysRevD.85.034503,Boucaud:2013jwa,Boucaud:2017obn,Boucaud2018}. Both facts
motivate the use of a phenomenologically modified gauge-fixed-Lagrangian with the addition of a gluon mass term \cite{Tissier:2010ts,Tissier:2011ey}.
This model is a phenomenological modification of Faddeev-Popov Lagrangian in Landau gauge which matches with Curci-Ferrari Lagrangian in the same gauge \cite{CF1976}.

The advantage of this phenomenological model is that it is renormalizable \cite{deBoer96} and IR safe, in the sense that it is possible to
find a renormalization scheme with no Landau pole \cite{Reinosa:2017qtf}. 
Within this model, one-loop calculations of the two and three point vertex functions have shown good agreement with lattice simulations
\cite{Tissier:2010ts,Tissier:2011ey,Pelaez:2013cpa}.
More recently a two-loop calculation have been performed for two-point correlation functions in quenched Landau-gauge Curci-Ferrari model \cite{Gracey:2019xom}, and again an overall
satisfactory description of the lattice data is obtained. Moreover, this model was also extended to finite temperature and chemical potential \cite{Reinosa:2015gxn,Reinosa:2014zta,Reinosa:2014ooa,Reinosa:2013twa}. 
In this work we present an improvement to the one-loop calculation of two-point correlation functions within the massive model presented in \cite{Tissier:2011ey}.
In previous works, as is common in the field, only one running coupling constant was 
considered. The logic behind this choice is that this is enough to renormalize the 
theory (in order to preserve gauge symmetry the interactions in different channels must 
be related to the same bare coupling constant). However, lattice simulations show that 
the strength of the interaction in different channels can be very different for momenta below 1 GeV (see e.g. \cite{Bogolubsky:2009dc,PhysRevD.85.034503,Boucaud:2013jwa,Athenodorou:2016oyh,Boucaud:2017obn,Boucaud2018,Zafeiropoulos2019}).
In order to take into account this effect, we propose to consider different 
renormalization factors for couplings extracted from different interactions.

The one-loop effect on the infrared behaviour of the different coupling constants in two-point correlation functions can also be obtained with a
two-loop calculation in the equal coupling case.
In this way, we can interpret our results as if some two-loops effects 
were taken into account. 
We stress that, even though these are two-loop corrections in the equal coupling case, the present
computation has still the simplicity of a one-loop calculation. We concentrate on the Landau gauge which has been largely studied in the past. 

The outline of the article is the following. In Section II
we describe the model and present our
one-loop calculation. We then describe in Section III
the IR safe renormalization scheme and introduce the renormalization group in Section IV. In Section V we show the comparison between our results
and lattice data, and discuss the relevance of treating differently the interactions.

\section{Model}

In order to study the IR regime of Yang-Mills theory we consider a model based on phenomenological observations.
We take as a starting point the Curci-Ferrari model in the Landau gauge \cite{CF1976}. 
It consists in adding a gluon mass term in Faddeev-Popov Lagrangian.
The addition of this term is motivated by lattice simulations that show a gluon propagator that saturates in the IR behaving as a massive propagator.
The advantage of this model is that it admits renormalization schemes without Landau pole and therefore perturbation theory is potentially under control even in the IR.
It is important to mention that the inclusion of a mass scale does not modify ultraviolet results since for momentum $p\gg m$ the mass scale can be neglected and, in this limit, the Faddeev-Popov Lagrangian is recovered.

In order to trace out the influence of different couplings in the Lagrangian,
we name the coupling constants for the three-gluons vertex, $g_A$, and for the ghost-gluon vertex, $g_C$. 
For simplicity, we identify the coupling extracted from the four-gluon vertex with the one associated with the three-gluon vertex.

In the Euclidean space the Landau-gauge Curci-Ferrari Lagrangian density reads: 

\begin{gather}
	\mathcal{L}=\dfrac{1}{4}F_{\mu\nu}^aF_{\mu\nu}^a+\partial_\mu\bar{c}^a(D_\mu c)^a+\dfrac{m^2}{2}A_\mu^aA_\mu^a+\textit{i}b^a \partial_\mu A_\mu^a, \nonumber
\end{gather}
where
\begin{gather}
	F_{\mu \nu}^a= \partial_\mu A_\nu^a-\partial_\nu A_\mu^a+g_A f^{abc}A_\mu^b A_\nu^c,\nonumber\\
	(D_\mu c)^a=\partial_\mu c^a + g_C f^{abc} A_\mu^b c^c.\nonumber
\end{gather}
The different labels in the couplings are in order to trace them out when renormalizing even though they are the same in the bare Lagrangian.

Accordingly, we depict the different vertices as follow:
\begin{figure}[h]
	\includegraphics[width=0.12\textwidth]{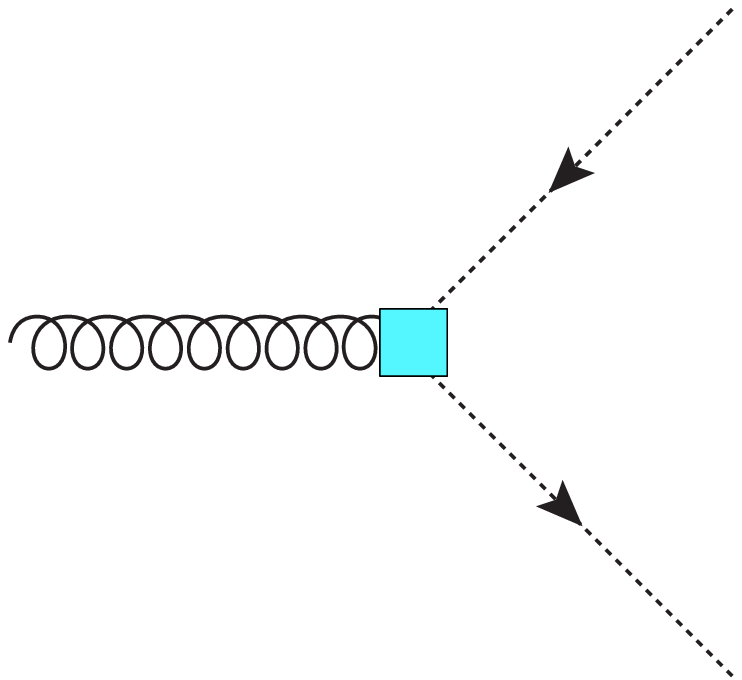}
	\includegraphics[width=0.11\textwidth]{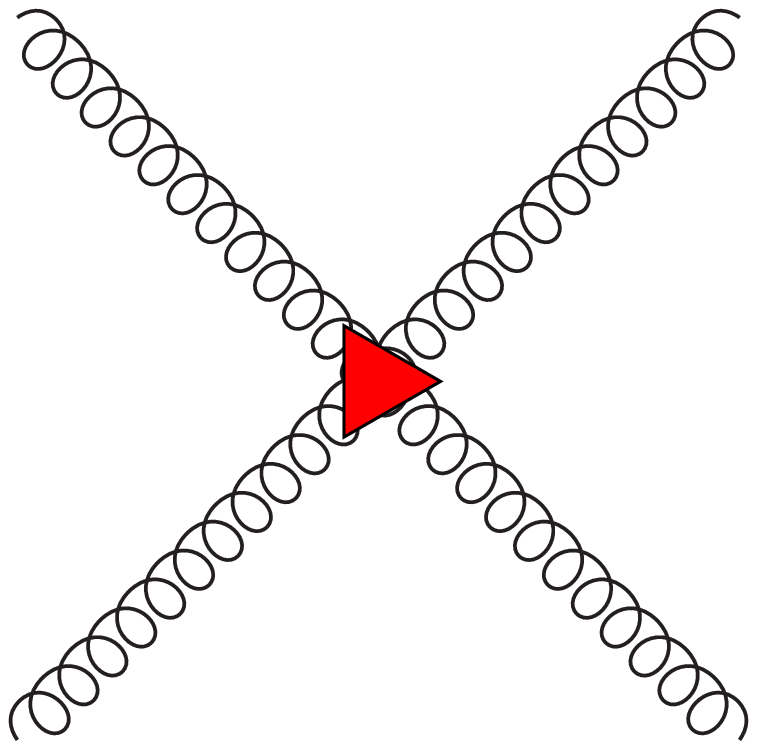}
	\includegraphics[width=0.12\textwidth]{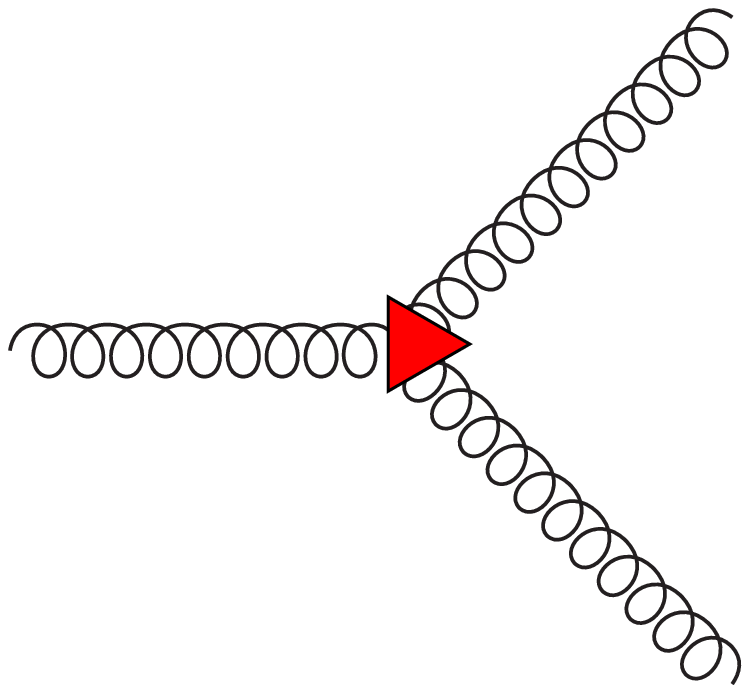}
\end{figure}

We will show that, even though coupling constants behave very differently in the infrared, this difference barely modifies the results of two-point functions.

\FloatBarrier

\subsection{Propagators}
In this section we compute one-loop correction for the gluon and ghost two-point vertex tracing out the influence of the different interactions.

The gluon two-point vertex function can be described by two scalar functions as
\begin{gather}
\Gamma_{\mu\nu}^{ab}(p)=\delta^{ab}\left(\Gamma^\perp(p)P_{\mu\nu}^\perp(p)+\Gamma^\parallel(p)P_{\mu\nu}^\parallel(p)\right),\nonumber
\end{gather}
where the projectors are defined as
\begin{gather}
 P_{\mu\nu}^\parallel(p)=\frac{p_\mu p_\nu}{p^2},\nonumber\\
 P_{\mu\nu}^\perp(p)=\delta_{\mu\nu}-P_{\mu\nu}^\parallel(p).\nonumber
\end{gather}

The diagrams contributing to $\Gamma_{\mu\nu}^{ab}(p)$  at one-loop are
\begin{figure}[h]
\vspace{-3mm}
	\includegraphics[width=0.15\textwidth]{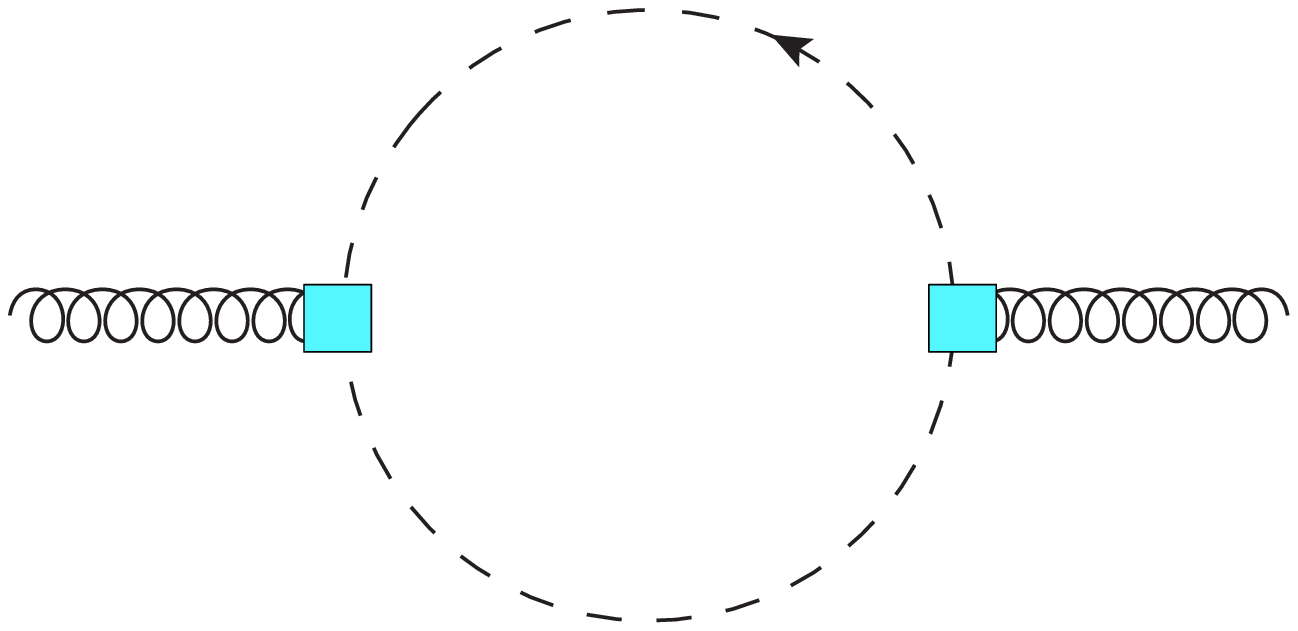}
	\includegraphics[width=0.15\textwidth]{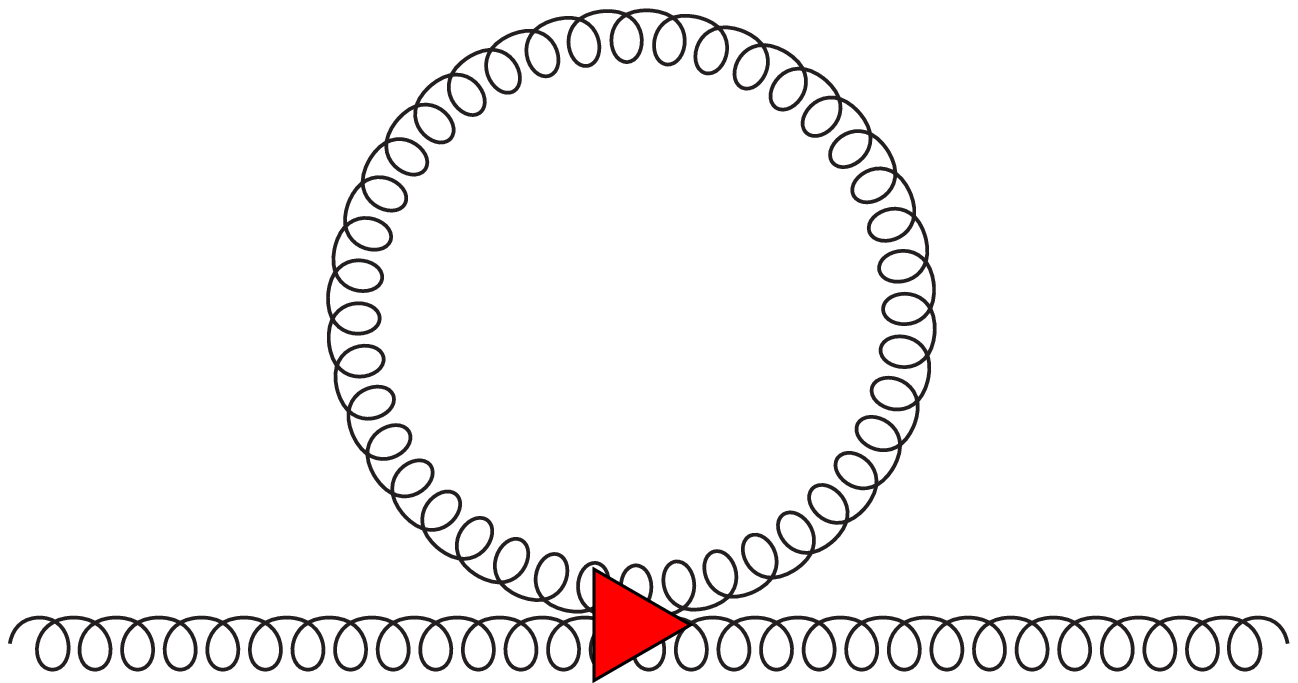}
	\includegraphics[width=0.15\textwidth]{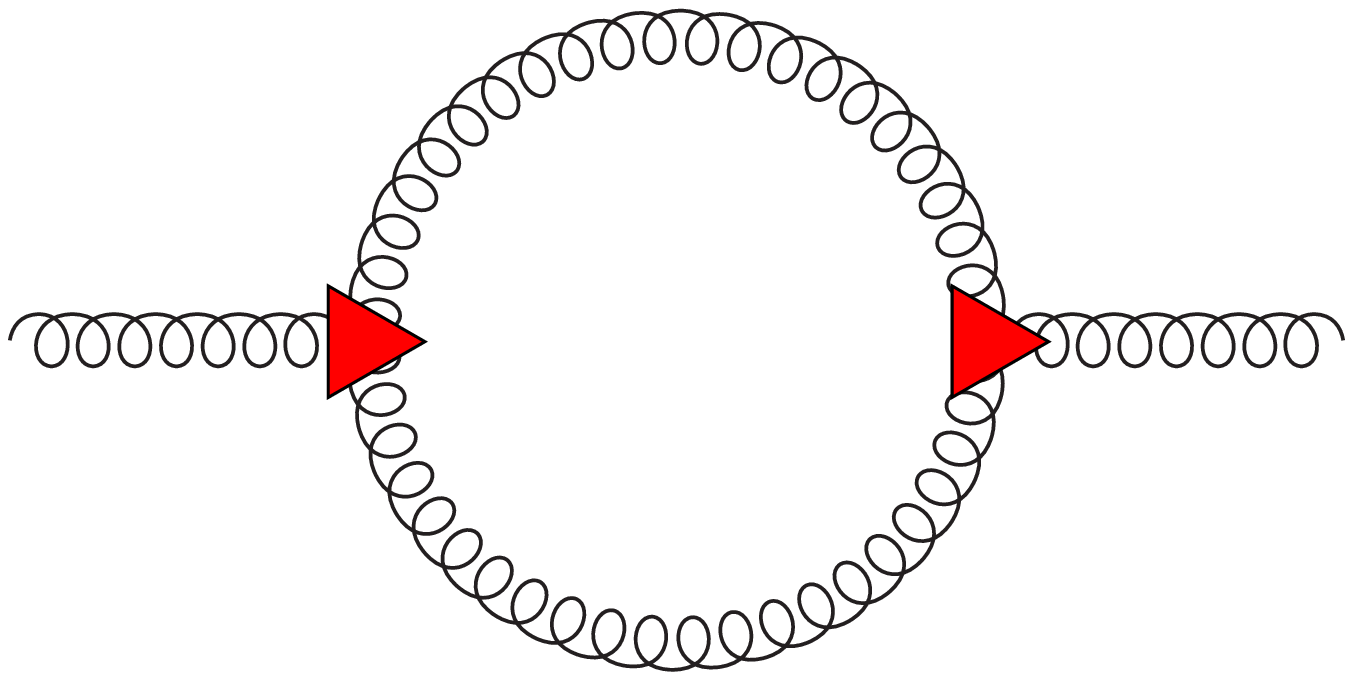}
	\caption{From left to right: $D_1$, $D_2$ and $D_3$.}
	\label{fig:1loopgluon}
\end{figure}
\FloatBarrier
The first two diagrams can be computed analytically for arbitrary dimension in terms of elementary functions \cite{Tissier:2011ey}.

\begin{gather}
\begin{aligned}
	D_1=&\dfrac{N (g_C^B)^2 \delta^{ab}}{(4\pi)^{d/2}}\dfrac{\Gamma(2-d/2)}{d-2}p^{d-4}\\
	& \times ((d-2)p_\mu p_\nu +p^2 \delta_{\mu\nu})\dfrac{\Gamma^2(d/2)}{\Gamma(d)},
\end{aligned}
\end{gather}

and

\begin{gather}
\begin{aligned}
	D_2=\dfrac{N(g_A^B)^2  \delta^{ab}}{(4\pi)^{d/2}}\dfrac{2(d-1)^2}{d(d-2)}m^{d-2}\Gamma(2-d/2)\delta^{\mu\nu}.
\end{aligned}
\end{gather}

For the last one, it is convenient to describe it in terms of one-loop master integrals 
\begin{gather}
\begin{aligned}
	& A(m)=\int\frac{d^dq}{(2\pi)^d}\frac{1}{q^2+m^2},\\
	& B(m_1,m_2)=\int\frac{d^dq}{(2\pi)^d}\frac{1}{q^2+m_1^2}\frac{1}{(q+p)^2+m_2^2},\nonumber
\end{aligned}
\end{gather}
that can be computed analytically for integer dimension in terms of elementary functions,
\begin{widetext}
\begin{gather}
\begin{aligned}
\small
	& D_3= (g_A^B)^2 N\delta^{ab} \left[ P_{\mu\nu}^\parallel(p) \left\{A(m)
   \left(d+\frac{1}{d}-\frac{m^2}{4 p^2}-\frac{7}{4}\right)-\frac{\left(m^2+p^2\right)^2 B(m,0)}{4
   p^2}\right\}\right.\\
   & \left.+ P_{\mu\nu}^\perp(p) \left\{-\frac{p^6 B(0,0)}{8 (d-1) m^4}-\frac{\left(4 m^2+p^2\right) B(m,m)
   \left(4 (d-1) m^4+4 (3-2 d) m^2 p^2+p^4\right)}{8 (d-1) m^4}\right.\right.\\
   &\left.\left.+\frac{\left(m^2+p^2\right)^2 B(m,0)
   \left(2 (3-2 d) m^2 p^2+m^4+p^4\right)}{4 (d-1) m^4 p^2}+\frac{A(m) \left(-4 d^2 p^4+d \left(m^4+5
   m^2 p^2+7 p^4\right)-4 m^2 p^2\right)}{4 (d-1) d m^2 p^2}\right\}\right].
\end{aligned}\nonumber
\end{gather}
\end{widetext}

The diagram contributing to the ghost two-point vertex function, $\Gamma_{c\bar c}(p)$, is
\begin{figure}[h]
	\includegraphics[width=0.25\textwidth]{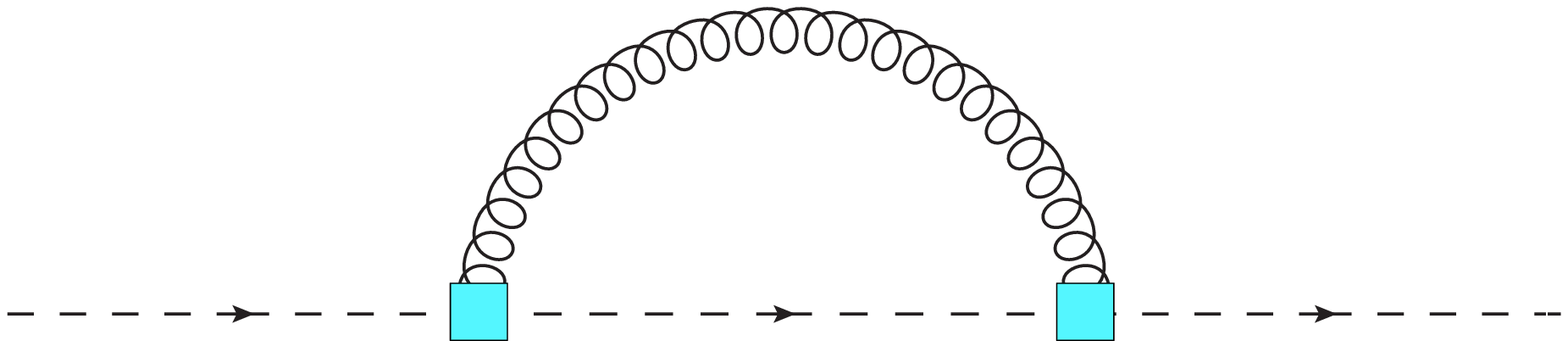}
\end{figure}
\FloatBarrier

\begin{equation}
\begin{split}
		\Gamma_{c^a \bar{c}^b}^{(2)}(p)=&\delta^{ab}p^2-\frac{\delta^{ab}N (g_C^B)^2}{4m^2}\Big((m^2-p^2)A(m)\\
		&-p^4B(0,0)+(m^2+p^2)^2B(0,m)\Big).\nonumber
\end{split}
\end{equation}

\section{Renormalization scheme}
In order to absorb the divergent part of one-loop diagrams, we introduce renormalization factors: 
\begin{gather}
\begin{aligned}
	& A_B^{a\,\mu}= \sqrt{Z_A} A^{a\,\mu},\hspace{1cm}	c_B^{a}= \sqrt{Z_c} c^{a},\\
	g^B_A= & Z_{g_A} g_A, \hspace{0.5cm} g^B_C= Z_{g_C} g_C,  \hspace{0.5cm}m_B^2= Z_{m^2} m^2.
\end{aligned}\label{eq:renfactors}
\end{gather}
The index $B$ represents the bare quantities and for now on, when
not specified, all quantities are the renormalized ones. 
In contrast with previous one-loop calculation for the equal coupling case, in this work we renormalize differently the ghost-gluon and the three-gluon couplings, 
even though they are related to the same bare value, $g^B_{A}=g^B_{C}$.

The field renormalization factors are defined in the scheme presented in \cite{Weber2014}, which is a reformulation of the IS-scheme of \cite{Tissier:2011ey}. 
In this renormalization scheme, renormalization factors for the fields and the gluon mass respect the conditions:
\begin{gather*}
\begin{aligned}
	&\Gamma^{\perp}_{A A}(\mu)= m^2+\mu^2, \\
 	&\Gamma_{c \bar{c} }(\mu)=\mu^2, \\
 	&\Gamma^{\parallel}_{A A}(\mu)=m^2.\\	
\end{aligned}\label{eq:renscheme}
\end{gather*}

The ghost-gluon coupling constant, $g_C$, is defined through the ghost-gluon vertex when the ghost momentum vanishes,
\begin{gather*}
\begin{aligned}
\Gamma_{A^a_\alpha C^b\bar{C}^c}^{(3)}(\mu,-\mu,0)	= 	-i f^{abc}  g_C \mu_\alpha \label{eq:gammac}.
\end{aligned}
\end{gather*}
As this vertex is finite, we have a non-renormalization theorem \cite{Taylor71} that allows to define the renormalization factor $Z{g_C}$ in terms of the fields renormalization factor as
\[Z_C \sqrt{Z_A} Z_{g_C}=1. \]
On the other hand, the three-gluon coupling is defined through the three-gluon vertex, $\Gamma_{A_\alpha^a\hspace{0.3mm} A_\beta^b  A_\gamma^c}^{(3)}(p,r,k)=-i f^{abc} g_A \Gamma_{A_\alpha\hspace{0.3mm} A_\beta  A_\gamma}(p,r,k)$. 
In particular, we define $g_A$ at the scale  $\mu$ so the vertex has a similar form as tree-level when one momentum is zero. 
The renormalization scheme can be summarized as:

\begin{gather*}
\begin{aligned}					
 	\Gamma_{A_\alpha^a\hspace{0.3mm} A_\beta^b  A_\gamma^c}^{(3)}(\mu,-\mu,0)
 	&=-i f^{abc}g_A \Gamma_{A_\alpha\hspace{0.3mm} A_\beta  A_\gamma}^{\text{tl}}(\mu,-\mu,0)
\end{aligned}
\end{gather*}
where  $\Gamma_{A_\alpha\hspace{0.3mm} A_\beta  A_\gamma}^{\text{tl}}(\mu,-\mu,0)$ represents the tensor structure of tree-level vertex with one vanishing momentum. In particular, 
\[\Gamma_{A_\alpha\hspace{0.3mm} A_\beta  A_\gamma}^{\text{tl}}(\mu,-\mu,0)=(\mu_\alpha\delta_{\beta\gamma}-2\mu_\beta\delta_{\alpha\gamma}+\mu_\gamma\delta_{\alpha\beta}).\]

\begin{figure}[h] 
	\centering
   	\includegraphics[width=0.8\linewidth]{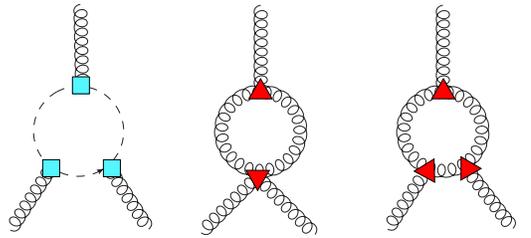}
   	\caption{Contribution at order one-loop of the vertex with three gluons.}
   	\label{fig:3vertexgluon1loop}
\end{figure}

One loop diagrams contributing to $\Gamma_{A_\alpha^a\hspace{0.3mm} A_\beta^b  A_\gamma^c}^{(3)}$ are depicted in Fig.\ref{fig:3vertexgluon1loop} and where computed in \cite{Pelaez:2013cpa}.
\FloatBarrier
For one momentum vanishing the vertex has only three scalar functions, $f_1(p^2)$, $f_2(p^2)$ and $f_3(p^2)$, defined as:

\begin{equation}
\begin{split}
			\Gamma_{A_\alpha^a\hspace{0.3mm} A_\beta^b \hspace{0.3mm}   A_\gamma^c}^{(3)\,B}&(p,-p,0)
			= f_1(p^2)(p_\gamma\delta_{\alpha\beta}+p_\alpha\delta_{\beta\gamma}) \\
			&+ f_2(p^2)p_\beta\delta_{\gamma\alpha}+f_3(p^2) p_\alpha p_\beta p_\gamma.
			\label{eq:Gammageneral}
\end{split}
\end{equation}

Using the relation $\Gamma_{A_\alpha^a\hspace{0.3mm} A_\beta^b \hspace{0.3mm} A_\gamma^c}^{(3)\,R}=Z_A^{\frac{3}{2}}\Gamma_{A_\alpha^a\hspace{0.3mm} A_\beta^b \hspace{0.3mm} A_\gamma^c}^{(3)\,B}$
 we obtain 
\begin{gather}
\begin{aligned}
	g_A&=\dfrac{-Z_A^{\frac{3}{2}}}{2}f_2(\mu^2,g_A^B,g_C^B). 
\end{aligned}\label{eq:defga}
\end{gather}

\subsection{The renormalization process with multiple couplings}
\label{sec.renormalizationproblems}

We observe that once the couplings related to different interactions in the 
Lagrangian, namely $g_A^B$ and $g_C^B$, are treated differently, the two-point vertex functions seems to have quadratic 
divergences.
This is coherent with the fact that we need gauge symmetry, and therefore it is necessary to respect $g_A^B=g_C^B$ 
in order to avoid this kind of divergence. This problem does not happen for the ghost vertex function at one-loop since it has only one kind of interaction.

However, in the gluon vertex function we find that the divergent parts are:

\begin{gather}
\begin{aligned}
\Gamma^\perp_\text{div}(p)&=\frac{N}{96 \pi^2 \epsilon}\left((g_C^B)^2p^2+(g_A^B)^2(25p^2-9m^2)\right),\nonumber\\
\Gamma^\parallel_\text{div}(p)&=\frac{N}{32 \pi^2 \epsilon}\left(\left((g_C^B)^2-(g_A^B)^2\right)p^2-3(g_A^B)^2m^2\right).
\end{aligned}
\end{gather}

Note that divergences in the parallel part proportional to $(g_C^B)^2-(g_A^B)^2$ spoil renormalization. 
In particular, a term of the form:
\begin{gather}
\label{problemaUV}
 \frac{Np^2\left((g_C^B)^2-(g_A^B)^2\right)}{348\pi^2}\left[\frac{12}{\epsilon}+6(\gamma+2)+6\text{Log}\left(\frac{p^2}{4\pi}\right)\right]
\end{gather}
where $\gamma$ is the Euler-gamma, appears in the bare parallel part of gluon vertex function. 

Obviously this kind of terms are zero since at bare level the coupling constants are the 
same. However, if we want the procedure to respect renormalizability even in perturbation theory, this would imply that the renormalized coupling constants at one-loop are the same (at one-loop 
all bare couplings can be replace by the renormalized ones). Thus, since we are mainly interested in describing phenomenologically the different behaviour of the coupling constants in the infrared with a 
simple calculation, we propose an ad-hoc procedure for this. Namely, we impose the equality of bare couplings in the ultraviolet divergences, but we still take into account the difference of the renormalized couplings in finite parts.

For practical purposes, in order to ensure the renormalization, we set term (\ref{problemaUV}) to zero but we still keep the various couplings different in the rest of the expression. 
This ad-hoc procedure allows to obtain a first study of the different behaviours of the coupling constants with a simple one-loop calculation.

In light of this, the divergent part of the renormalization factors in four dimensions, using  $d= 4-\epsilon $, are given by
\begin{gather}
\begin{aligned}
     Z_C&= 1 + \dfrac{3 g_C^2 N}{32 \epsilon \pi^2}+\mathcal{O}(g^4),\\
     Z_{m^2}&= 1-\dfrac{34 g_A^2 N + g_C^2 N}{96 \epsilon \pi^2}+\mathcal{O}(g^4),\\
     Z_A&= 1+\dfrac{25 g_A^2 N + g_C^2 N}{96 \pi^2 \epsilon}+\mathcal{O}(g^4).
     \label{eq:divrenfact}
\end{aligned}
\end{gather}

\section{Renormalization group}

Once we find the renormalization factors we can compute the $\beta$ and $\gamma$ functions, defined as:
\begin{equation}
\begin{gathered}
	\gamma_A=\mu \dfrac{\partial \log(Z_A)}{\partial \mu},\\
	\gamma_C=\mu \dfrac{\partial \log(Z_C)}{\partial \mu},\\
	\beta_{m^2}=\mu\dfrac{\partial m^2}{\partial \mu}\Big|_{g_C^B,g_A^B, m^2_B}=-m^2\gamma_{m^2},\\
	\beta_{g_A}=\mu\dfrac{\partial g_A}{\partial \mu}\Big|_{g_C^B,g_A^B, m^2_B}=\left(\dfrac{3}{2}\gamma_A+\mu \dfrac{\mathrm{d}\log(B(\mu^2))}{\mathrm{d}\mu}\right)g_A,\\
	\beta_{g_C}=\mu\dfrac{\partial g_C}{\partial \mu}\Big|_{g_C^B,g_A^B, m^2_B}=\left(\gamma_C+\dfrac{1}{2}\gamma_A\right)g_C.
	\label{eq:rensystofeq}
\end{gathered}
\end{equation}

Once we solved this system of differential equations we can find the flow for the different coupling constants and masses.\\
The $\beta$ functions in the UV for $g_A$ and $g_C$, keeping track of the couplings, are
\begin{eqnarray}
\beta_{g_A}^{UV}(g_C,g_A)&=&\dfrac{N}{192 \pi^2}(-42g_A^3-3g_Ag_C^2+g_C^3)+\mathcal{O}(g^5), \nonumber\\
\beta_{g_C}^{UV}(g_C,g_A)&=&-\dfrac{N g_C}{192 \pi^2}(25g_A^2+19 g_C^2)+\mathcal{O}(g^5). \nonumber
\end{eqnarray}
These expressions match with standard one-loop $\beta$-function, $\beta_g=-\frac{11 N g^3}{3(16 \pi^2)}$, when $g_A=g_C$.
In order to obtain the propagators renormalized at a different scale we use the renormalization group equation:
\begin{equation}
\begin{split}
	\Big( \mu\dfrac{\partial}{\partial\mu}+&\dfrac{1}{2}(n_A\gamma_A+n_C\gamma_C)+\beta_{g_A}\dfrac{\partial}{\partial g_A}\\
	 &+\beta_{g_C}\dfrac{\partial}{\partial g_C}+\beta_{{m^2}}\dfrac{\partial}{\partial {m^2}}\Big)\Gamma^{(n_A,n_C)}_{R}=0,
\end{split}
\end{equation}
which is a generalization of the standard renormalization group equation for this model.

The vertex function with different energies scales are related as
\begin{equation}
\begin{split}
\Gamma^{(n)}_{R}\left(\{p\},\mu,g_A,\right.&\left.g_C\right)=z_A(\mu)^{\frac{n_A}{2}}z_C(\mu)^{\frac{n_C}{2}}\\
&\times \Gamma^{(n)}_{R}\left(\{p\},\mu_0,g_A,g_C,\right).
\label{eq:gammamugammamu0}
\end{split}
\end{equation}

In equation \eqref{eq:gammamugammamu0} we took the energy $\mu$ as the reference point in order to resolve the RG equation, with $z_A$ and $z_C$ defined as
\begin{gather}
\begin{aligned}
\mathrm{log}\left(z_A(p)\right)=\int_{\mu}^{p} \dfrac{\mathrm{d}\mu'}{\mu'}\gamma_A\left(g_A(\mu'),g_C(\mu'),m^2(\mu')\right),\\
\mathrm{log}\left(z_C(p)\right)=\int_{\mu}^{p} \dfrac{\mathrm{d}\mu'}{\mu'}\gamma_C\left(g_A(\mu'),g_C(\mu'),m^2(\mu')\right).
\label{eq:logdez}
\end{aligned}
\end{gather}

Therefore two-point functions are computed as:
\begin{gather}
\begin{aligned}
\Gamma_{AA}^{(2)}(p,\mu_0,g_A,g_C,m^2)=\frac{p^2+m^2(p^2)}{z_A(p)}=\frac{p^2}{D(p)},\\
\Gamma_{c\bar{c}}^{(2)}(p,\mu_0,g_A,g_C,m^2)=\frac{p^2}{z_C(p)}=\frac{p^2}{J(p)}
\label{eq:props}
\end{aligned}
\end{gather}
where $D(p)$ and $J(p)$ denote the gluon and ghost dressing functions respectively.

\section{Results}

In order to obtain the behavior of the coupling constants and consequently the propagators, we have to give the initial condition of the renormalization group flow. 
In this case we choose the initial value at $\mu_0=10$ GeV.  
A priori we can think that we have three free parameters to fit, $g_A(\mu_0)$, $g_C(\mu_0)$ and $m(\mu_0)$. 
However, as both coupling constants have the same bare value, we can relate them at large energy ($\mu_0\approx 10$ GeV) using standard perturbation theory. We specify the calculation of this relation in appendix \ref{sec.relgagcUV}. 
Perturbativelly  $g_A$ and $g_C$ can be related as,
\begin{equation}
	g_A=g_C+ \dfrac{37N}{384 \pi^2}g_C^3+ \mathcal{O}\left(g_C^5\right).
	\label{eq:relaciongagc}
\end{equation}
For large energy $g_A$ is barely larger that $g_C$.
Using this relation, at the end, we have to fix only $m_0=m(10 \, \textnormal{GeV})$ and $g_{C,0}=g_C(10 \, \textnormal{GeV})$ in order to fit both propagators at the same time. 
The parameters are chosen in order to
 minimize the error function $\chi_{\mathrm{tot}}$ defined as 
\begin{eqnarray}
	\chi_{\mathrm{tot}}^2=\dfrac{\chi_{\mathrm{abs}}^2+\chi_{\mathrm{rel}}^2}{2}.
\end{eqnarray} 
 
\begin{figure}[h]
   	\centering
      \includegraphics[width=\linewidth]{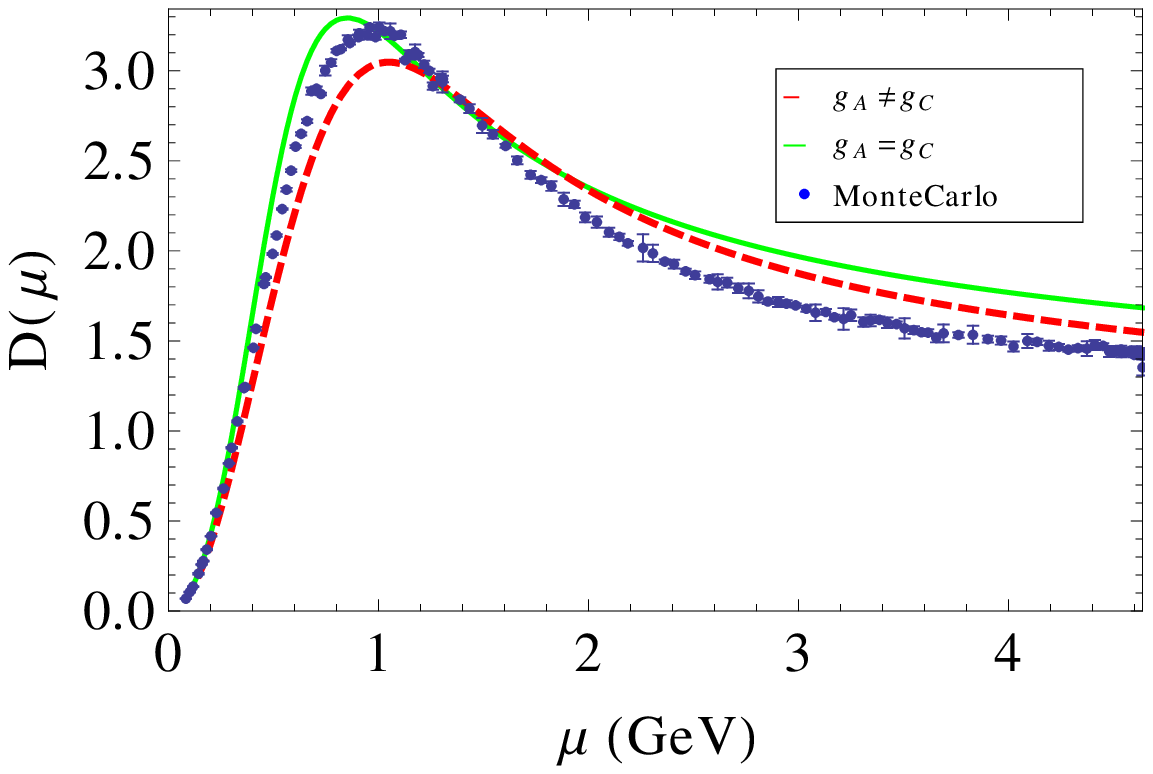}
    \includegraphics[width=\linewidth]{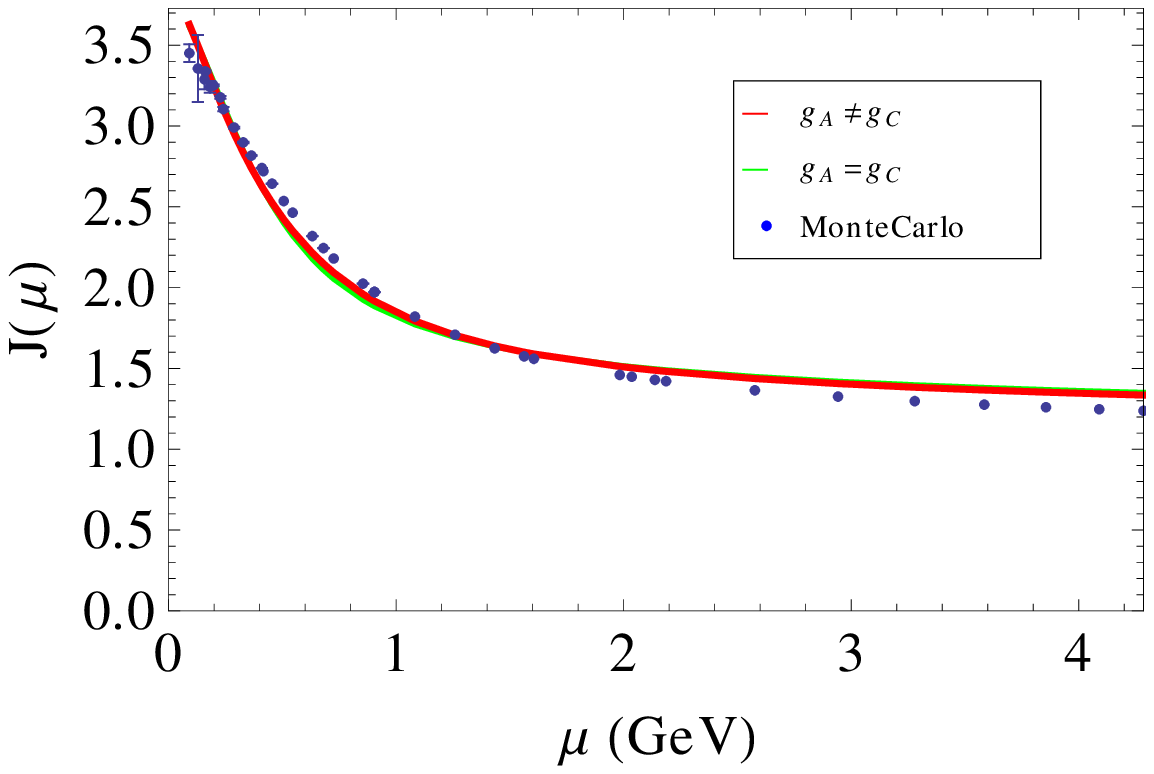}
   \caption{Gluon (top) and ghost (bottom) dressing function as a function of the scale $\protect\mu$ computed with different IR couplings (dashed) and in the equal renormalized coupling case (full). Lattice data from \cite{Bogolubsky:2009dc}}
\label{fig:resultadosjuntosghost}
\end{figure}

Relative and absolute error are defined as follow:

\begin{eqnarray}
	\chi_{\mathrm{rel}}^2&=&\dfrac{1}{ M} \sum_{i=1}^{ M}\dfrac{({y_l}_{i}-\mathrm{A}\times{y_t}_{i})^2}{{y_l}_{i}^2},\nonumber\\
	\chi_{\mathrm{abs}}^2&=& \dfrac{1}{ M y_l(1GeV)^2}\sum_{i=1}^{M}({y_l}_{i}-\mathrm{A}\times{y_t}_{i})^2,\nonumber
\end{eqnarray}
where $M$ is the number of lattice data and A is the multiplicative factor that appears due to the difference between the IS-renormalization scheme and lattice renormalization. It takes the form
\begin{eqnarray}
	\mathrm{A}&=&\dfrac{\sum_{i=1}^{M}{y_l}_{i} {y_t}_{i}}{\sum_{i=1}^{M}{y_t}_{i}^2},\nonumber
\end{eqnarray}
in order to minimize the quadratic error.
The parameters that minimize $\chi$ are  $g_{C,0}=1.67$, $m_0=0.24$ GeV. In Fig. \ref{fig:resultadosjuntosghost} we show the comparison of the gluon
propagator of this work, computed with $g_{C,0}=1.67$, $m_0=0.24$ GeV, to the lattice data in $SU(3)$ from \cite{Bogolubsky:2009dc}.

\begin{figure}[h]
   \centering
   \includegraphics[width=\linewidth]{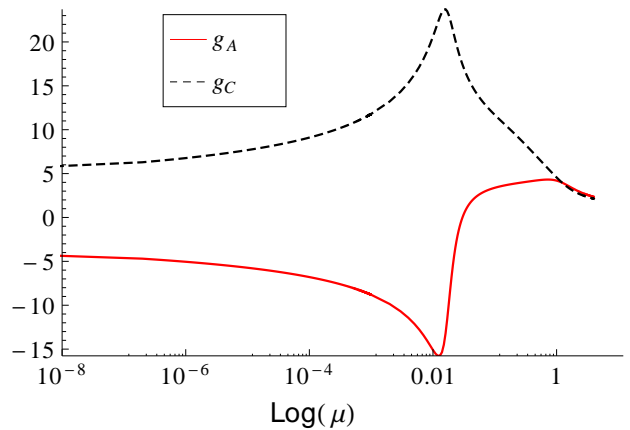}
   \caption{Flow of the coupling constants $g_A$ (full, red online) and $g_C$ (dashed, black online) as a function of Log($\mu$).}
   \label{fig:flowgagc}
\end{figure}
From Fig. \ref{fig:resultadosjuntosghost} we infer that a good fitting is obtained from this model. The fitting is only a bit better than in the equal coupling case \cite{Tissier:2011ey}. 
The fact that there is no great difference with the equal coupling case is consistent with the fact that the perturbative expansion is under control.

Apart from the study of two-point functions we are able to study in detail the infrared behaviour of the coupling constants. With this purpose, we study the low momenta limit $\mu\ll m$ of Eqs. (\ref{eq:rensystofeq}). Therefore, the one-loop $\beta$ functions in the infrared regime read,
\begin{eqnarray}
	\beta_{g_C}^{IR}(g_C,g_A)&=&\dfrac{N}{192 \pi^2}(3g_A^2g_C-g_C^3)+\mathcal{O}(g^5),\nonumber\\
	\beta_{g_A}^{IR}(g_C,g_A)&=&\dfrac{N}{192 \pi^2}(9g_A^3 -3g_Ag_C^2+g_C^3)+\mathcal{O}(g^5),\nonumber\\
	\beta_{m^2}^{IR}(g_C,g_A)&=&\dfrac{N m^2}{96 \pi^2}  (3g_A^2-g_C^2)+\mathcal{O}(g^5).
	\label{eq:betaenelIR}
\end{eqnarray}
The analytic solution of the system \eqref{eq:betaenelIR} shows that both coupling constants go to zero in the very deep IR, see appendix \ref{sec.gAgCdeepIR}. 
While the ghost coupling goes to zero by positives values, the gluon coupling arrives from negatives values. 
The zero crossing is also observed by lattice simulations \cite{Duarte:2016ieu,Athenodorou:2016oyh,Boucaud:2017obn}. This is due to the fact that the three-gluon vertex 
in the infrared is dominated by massless ghost while gluons are frozen \cite{Pelaez:2013cpa,Aguilar:2019jsj}.

\begin{figure}[h]
   \centering
   \includegraphics[width=\linewidth]{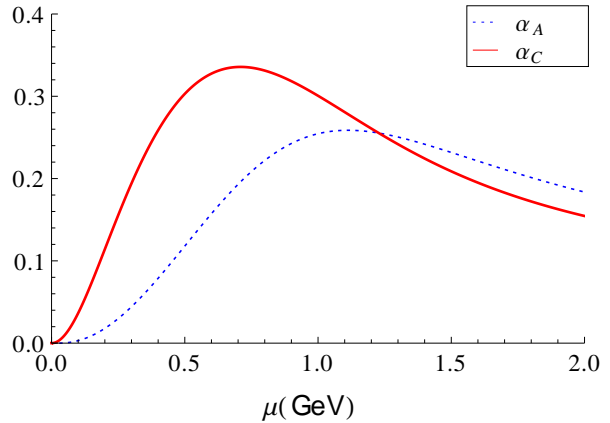}
   \caption{Comparison of various expansion parameters.}
   \label{fig:todosalphas}
\end{figure}

Still we can wonder why even though coupling constants are extremely different in the infrared this has almost no effects in the results for two-point functions. 
We think that it is because they start differing
below 1 GeV as it is shown in Fig. \ref{fig:flowgagc}. 
Moreover, at one-loop, the expansion parameter related to the ghost-gluon coupling is in fact $\alpha_C^{\tiny\text{1-loop}}=\frac{N g_C^2}{16\pi^2}$
while the expansion parameter of the three-gluon coupling is $\alpha_A=\frac{N g_A^2}{16\pi^2}\frac{\mu^2}{\mu^2+4m^2}$. In the last case, the extra factor appears due to massive gluons and to the fact that, since we are coupling 1PI functions, the internal gluons come at least in pairs. 
The three-gluon expansion parameter is suppressed in the infrared and therefore is almost equivalent to consider $g_A$ or $g_C$ in that quantity.
Let us note that for higher loops the expansion parameters is for both couplings $\alpha_{A,C}=\frac{N g_{A,C}^2}{16\pi^2}\frac{\mu^2}{\mu^2+4m^2}$, since higher loop diagrams have internal gluons. 
The latter is less than $0.3$ as it can be appreciated in Fig. \ref{fig:todosalphas}.  

\section{Conclusion}

In this work we present a one-loop calculation of two-point
correlation functions using a modified Lagrangian in Landau gauge. Based on the fact that coupling constants related to different vertices behave differently in the infrared, we do not identify vertices with only one coupling as presented in \cite{Tissier:2011ey}.

As it was mentioned, this one-loop difference on the couplings can be also taken into account in a two-loop calculation using the model with only one coupling constant.
Therefore, the fact that this results are of the same quality as the equal-coupling case show that perturbation theory seems to be under control. 
A complete two loop calculation of these quantities has been recently reported \cite{Gracey:2019xom}.

We can appreciate in Fig. \ref{fig:flowgagc} the change of sign in the gluon coupling constant, $g_A$, associated to the three gluon vertex. 
The change of sign is due to the contribution of the diagram with one-loop of ghosts (left of Fig. \ref{fig:3vertexgluon1loop}). 
We find that, even though both coupling constants differ considerably in the infrared, the gluon and ghost propagators slightly improve with respect to the one coupling case of \cite{Tissier:2011ey}.
This can be understood since the expansion parameters differ below the gluon mass scale, and in that regime gluon fluctuations are almost frozen.

Even though we do not know how to justify this massive Lagrangian from first principles, we think that the simplicity of this model makes it useful to study phenomenologically different IR properties of Yang-Mills theory.

\begin{acknowledgments}
We are very thankful to Nicolás Wschebor, Matthieu Tissier and Emerson Luna for useful discussions and reading previous version of this manuscript.
We also thank the support of ECOS program, ANII ``$FCE\_126412$'' research project and the support of a CNRS-PICS project “irQCD”. M. Fernández wants to thank the Comisión Académica de Posgrado de la Universidad de la República for financial support.

\end{acknowledgments}

\appendix

\section{Relation between $g_A$ and $g_C$ in the ultraviolet}
\label{sec.relgagcUV}

For $\mu\gg m$, the calculation is identical to the one in standard Yang-Mills. We can use perturbation theory, in this way we can relate the renormalize couplings with the bare ones as follows: 
\begin{equation}
	\begin{cases}
			g_A=& g_B+ f_A\left(\frac{\mu}{\Lambda} \right) g_B^3+\mathcal{O}(g_B^5),\\ 
			g_C=& g_B+f_C\left(\frac{\mu}{\Lambda} \right)g_B^3+\mathcal{O}(g_B^5)
	\end{cases}
\end{equation}
were $\Lambda$ is the UV scale. At one-loop order we can write:
\begin{equation}\label{eq:gagc}
	\begin{cases}
		g_A=g_B+ f_A\left(\frac{\mu}{\Lambda} \right) g_C^3+\mathcal{O}(g_C^5),\\
		g_C= g_B+f_C\left(\frac{\mu}{\Lambda} \right)g_C^3+\mathcal{O}(g_C^5).
	\end{cases}
\end{equation}
From this we conclude:
\begin{equation}
	\begin{cases}
		g_B=g_C-f_C \left(\frac{\mu}{\Lambda} \right)g_C^3+\mathcal{O}(g_C^5), \\
		g_A=g_C+ \left( f_A\left(\frac{\mu}{\Lambda} \right)- f_C\left(\frac{\mu}{\Lambda} \right) \right)g_C^3+\mathcal{O}(g_C^5).
	\end{cases}
\end{equation}
Because $g_A$ and $g_C$ are renormalized couplings both have a limit for $\Lambda\rightarrow\infty$. 
This implies that $f_A \left(\frac{\mu}{\Lambda} \right)- f_C \left(\frac{\mu}{\Lambda}\right)$ has a limit for $\Lambda\rightarrow\infty$. 
For dimensional reasons, this subtraction must be a pure number.
The corresponding result for $f_A$ and $f_C$ is extracted from the definition of $g_A$ y $g_C$ using the one-loop vertices computed in \cite{Pelaez:2013cpa}.

\section{Coupling constants in the deep IR}
\label{sec.gAgCdeepIR}

The IR regime is characterized by energies that satisfy $\mu\ll m$, being $\mu$ the energy and $m$ the gluon mass. The functions $\beta$ in this regime are,

\begin{eqnarray}
	\beta_{g_C}^{IR}(g_C,g_A)&=&\dfrac{N}{192 \pi^2}(3g_A^2g_C-g_C^3)+\mathcal{O}(g^5),\nonumber\\
	\beta_{g_A}^{IR}(g_C,g_A)&=&\dfrac{N}{192 \pi^2}(9g_A^3 -3g_Ag_C^2+g_C^3)+\mathcal{O}(g^5),\nonumber\\
	\beta_{m^2}^{IR}(g_C,g_A)&=&\dfrac{N m^2}{96 \pi^2}  (3g_A^2-g_C^2)+\mathcal{O}(g^5).
\end{eqnarray}
We rewrite these equations, at order $g^3$, using the changes of variables: $t=\mathrm{log}\left(\frac{\mu}{\bar{\mu}}\right)$ where $\bar{\mu}$ is an arbitrary constant and $y=\frac{g_C^3}{g_A}$. 
\begin{gather}
\begin{aligned}
	\dfrac{1}{g^2_C}\dfrac{\mathrm{d} g_C^2}{\mathrm{d} t}&= \dfrac{1}{m^2}\dfrac{\mathrm{d} m^2}{\mathrm{d} t}\\
	\dfrac{\mathrm{d} g_C}{\mathrm{d} t}&=\dfrac{g_C}{64 \pi^2}(3g_A^2-g_C^2)\\
	\dfrac{\mathrm{d} }{\mathrm{d} t}\left(\dfrac{1}{y}\right)&= \dfrac{1}{(8\pi)^2} 
\end{aligned}
\label{eq:sistemadeecgagcIR}
\end{gather}
Using the last equation of system \eqn{eq:sistemadeecgagcIR} we get
\begin{equation}
	g_A=\dfrac{g_C^3 t}{(8\pi)^2}.
	\label{eq:vinculogagcIR}
\end{equation}
Substituting the relation \eqn{eq:vinculogagcIR} in the second equation of the system \eqn{eq:sistemadeecgagcIR} we will have a non autonomous equation:
\begin{gather}
	\begin{aligned}
	\dfrac{\mathrm{d} g_C}{\mathrm{d} t}&=\dfrac{g_C^3}{(8\pi)^2}\left(\dfrac{3g_C^4 t^2}{(8\pi)^4}-1\right).
	\label{eq:ecuacionnohomogenea}
	\end{aligned}
\end{gather}
As we are interested in scales $\mu\ll m$, $t$ will be negative, $t<0$. 
We transform Eq.\eqn{eq:ecuacionnohomogenea} into an autonomous equation using $\lambda=g_C (-t)^{1/2} $ and $x=\mathrm{log}|t|$, 
\begin{equation}
	\partial_x \lambda= \lambda\left(-\dfrac{1}{2}+\dfrac{3\lambda^6}{(8\pi)^6}-\dfrac{\lambda^2}{(8\pi)^2}\right).
	\label{eq:ecdifgagcIRhomogenea}
\end{equation}

Because the differential equation \eqn{eq:ecdifgagcIRhomogenea} is homogeneous in $x$, when $x\rightarrow\infty$, that is, for small energies, $\lambda\rightarrow$ constant. 
We could find the fixed points of $\lambda$ imposing $\partial_x \lambda=0$.
This is equivalent to solve 
\begin{gather}
\begin{aligned}
	\dfrac{1}{2}+\sigma-3\sigma^3=0.
\end{aligned}
\end{gather}
with $\sigma=\frac{\lambda^2}{(8 \pi)^2}$.
We can find numerically that the only real and positive root is $\sigma_0\sim 21.709$,
\begin{gather}
\begin{aligned}
	g_C^{IR}&=\frac{\sigma_0}{\sqrt{-t}},\\
	g_A^{IR}=\frac{g_C^3 t}{64 \pi^2}&=\dfrac{-\sigma_0^3}{64\pi^2\sqrt{-t}}.
	\label{eq:valoresgaygcenelIR}
\end{aligned}		
\end{gather}

The analytic result seen in the system \eqref{eq:valoresgaygcenelIR} shows us that both coupling constants goes slowly to zero when the energy decrease. The ghost coupling constant does this by the positives values and the gluon one by the negatives.
We know that the latter sign comes from ghost loop of the three-gluon vertex. \\
As a verification we can observe when we impose the approximation of $g_C\approx g_A$ at first order we get the beta function $\beta_{g_c}$:
\begin{eqnarray}
	\beta_{g_c}=\dfrac{g^3N}{96\pi^2}
\end{eqnarray}
This match with the infrared limit of the beta function, $\beta_g$, calculated for the infrared scheme in \cite{Pelaez2015bqb}.

\newpage
\addcontentsline{toc}{section}{Bibliografía} 
\bibliography{library}

\end{document}